\documentclass[pra,aps,reprint,amsmath,superscriptaddress,showkeys]{revtex4-1}

\usepackage{bm}
\usepackage{graphicx}
\usepackage{dsfont}
\usepackage{xcolor}

\newcommand{\bra}[1]{\langle #1 | \,}
\newcommand{\ket}[1]{\, | #1 \rangle}
\newcommand{\braket}[2]{\langle #1 | #2 \rangle}
\newcommand{\expv}[1]{\langle #1 \rangle}

\newcommand{\mr}[1]{\mathrm{#1}}
\newcommand{\mc}[1]{\mathcal{#1}}

\newcommand{\om}{\omega}

\newcommand{\eps}{\epsilon}
\newcommand{\veps}{\varepsilon}

\begin{document}

\title{Optimal collection of radiation emitted by a trapped atomic ensemble}

\author{\'A. Kurk\'o}
\affiliation{Wigner Research Centre for Physics, H-1525 Budapest, P.O. Box 49., Hungary}

\author{P. Domokos}
\affiliation{Wigner Research Centre for Physics, H-1525 Budapest, P.O. Box 49., Hungary}

\author{A. Vukics}
\affiliation{Wigner Research Centre for Physics, H-1525 Budapest, P.O. Box 49., Hungary}

\author{T. B\ae kkegaard}
\affiliation{Department of Physics and Astronomy, Aarhus University, DK-8000 Aarhus C, Denmark}

\author{N. T. Zinner}
\affiliation{Department of Physics and Astronomy, Aarhus University, DK-8000 Aarhus C, Denmark}

\author{J. Fort\'agh}
\affiliation{Physikalisches Institut, Eberhard Karls Universit\"at T\"ubingen, 
D-72076 T\"ubingen, Germany}

\author{D. Petrosyan}
\affiliation{Physikalisches Institut, Eberhard Karls Universit\"at T\"ubingen, 
D-72076 T\"ubingen, Germany}
\affiliation{Institute of Electronic Structure and Laser, Foundation for Research and Technology -- Hellas, 
GR-70013 Heraklion, Crete, Greece}

\date{\today}

\begin{abstract}
Trapped atomic ensembles are convenient systems for quantum information storage in the long-lived 
sublevels of the electronic ground state and its conversion to propagating optical photons via
stimulated Raman processes.  
Here we investigate a phase-matched emission of photons from a coherently prepared atomic ensemble. 
We consider an ensemble of cold atoms in an elongated harmonic trap with normal density distribution, 
and determine the parameters of paraxial optics to match the mode geometry of the emitted 
radiation and optimally collect it into an optical waveguide.
\end{abstract}


\maketitle

\section{Introduction}
An important yet difficult task in quantum information and communication is  
deterministic generation of optical photons in well-defined spatial and temporal modes. 
Photons can serve as flying qubits to encode quantum information and reliably transmit 
it over long-distances via quantum channels \cite{Kimble2008,OBrien2009}. 
For optical photons, such quantum channels are free space or fiber waveguides. 
On the other hand, atomic ensembles have good coherence properties and strong dipole transitions
for efficient coupling to optical photons \cite{EITrev2005,Hammerer2010,Sangouard2011}.
Moreover, atoms can couple to microwave fields and thereby be interfaced with superconducting circuits, 
which are among the most advanced quantum processors \cite{Kurizki2015}. 
The atomic ensembles can then play the role of the quantum memories and microwave to optical transducers 
\cite{Kurizki2015,Petrosyan2019,Covey2019} to interconnect distant superconducting quantum circuits via optical fibers. 

Here we consider photon emission from an atomic ensemble coherently prepared in a collective storage 
state corresponding to a single symmetric spin-wave excitation. This preparation step may be the result of
collective coupling of alkali atoms on the ground state hyperfine transition \cite{Verdu2009,Hattermann2017} 
or on the highly excited Rydberg transition \cite{Petrosyan2009,Hogan2012} to a microwave resonator that in turn 
contains superconducting qubits. A spontaneous Raman process triggered by a coupling laser pulse converts 
the collective spin excitation of the atoms to an optical photon emitted predominantly in the phase-matched direction. 
We examine the spatio-temporal profile of the emitted photon and determine the parameters of the paraxial optical elements 
to optimally collect this photon into a single-transverse-mode optical waveguide. 
The waveguide can then transmit the photon encoding the quantum information to a distant atomic ensemble, where a reverse process, 
equivalent to dynamical light storage in an electromagnetically induced transparency (EIT) medium \cite{EITrev2005}, 
will coherently convert the optical excitation to a collective microwave spin excitation.

The paper is organized as follows. 
In Sec. \ref{sec:atphot} and in Appendices \ref{app:sphwf} and \ref{app:satphot}, we present the mathematical 
formalism to describe the photon emission by coherently prepared atoms and photon collection by Gaussian optics. 
In Sec. \ref{sec:results} we present the results of analytic and numerical calculations for optimal photon collection, 
followed by conclusions in Sec.~\ref{sec:conclud}.

\section{Interaction of atoms with a radiation field}
\label{sec:atphot}

%%%%%%%%%%%%%%%%%%%%%%%%%%%%%%%%%%%%
\begin{figure}[b]
\centering
\includegraphics[width=1.0\linewidth]{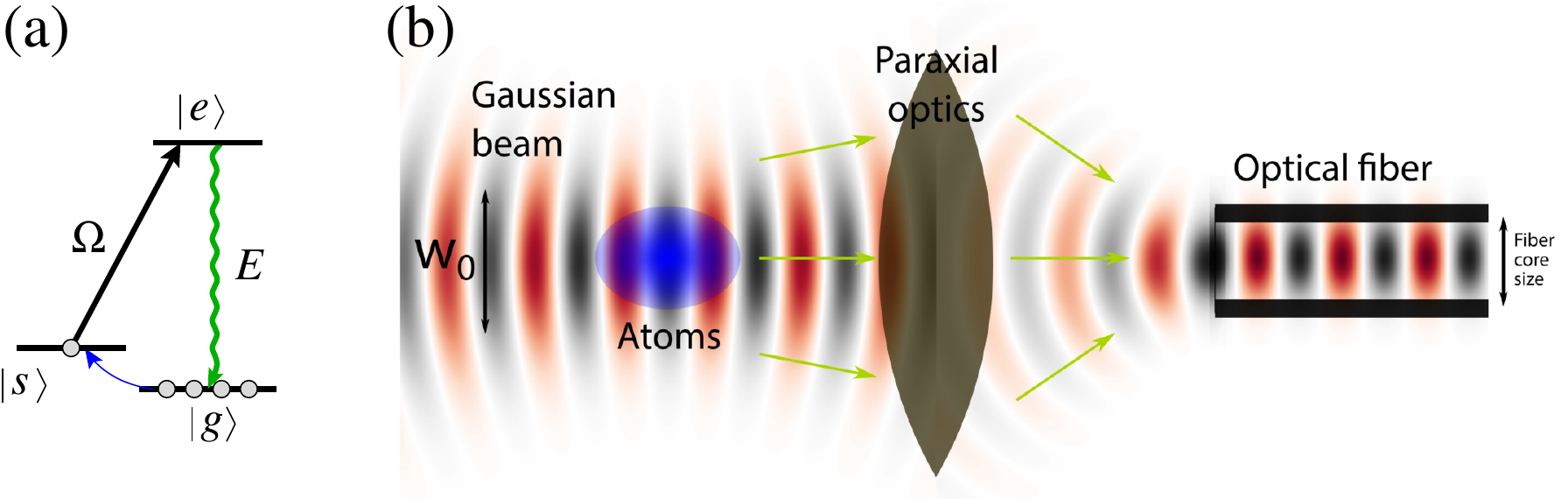}
\caption{(a) Level scheme of three-level atoms. With all the atoms initially in the ground state $\ket{g}$,
a weak microwave or Raman transition (blue arrow) creates a single collective spin excitation in the storage state $\ket{s}$.
A laser pulse acting on the $\ket{s} \to \ket{e}$ transition with Rabi frequency $\Omega$ then converts the collective spin
excitation to a single-photon field $E$ emitted on the $\ket{e} \to \ket{g}$ transition. 
(b) The photon is emitted predominantly in the phase matched direction into a Gaussian mode 
with waist $w_0$ and is collected into a fiber waveguide by paraxial optical elements.}
\label{fig:Scheme}
\end{figure}
%%%%%%%%%%%%%%%%%%%%%%%%%%%%%%%%%%%%

Consider an ensemble of $N$ three-level atoms with the long-lived ground-state sublevels $\ket{g}$ and $\ket{s}$ 
and an electronically excited level $\ket{e}$, as shown in Fig.~\ref{fig:Scheme}(a). 
We assume that the atoms initially in the ground state $\ket{G} \equiv \ket{g_1,g_2,\ldots,g_N}$ 
are transferred to the collective single-excitation storage state 
$\ket{S} = \frac{1}{\sqrt{N}} \sum_{j=1}^N \ket{g_1,g_2,\ldots,s_j, \ldots, g_N}$ 
by a weak microwave or Raman process that acts symmetrically on all the atoms. 
A spatially uniform laser pulse couples near-resonantly the storage state $\ket{s}$ 
to the excited state $\ket{e}$ with Rabi frequency $\Omega$. An atom at position
$\bm{r}$ in the excited state $\ket{e}$ then emits a photon into the free-space radiation field 
$\hat{E}(\bm{r}) = \sum_k \hat{a}_k u_k(\bm{r})$, with modes $\{u_k (\bm{r}) \}$ forming a complete basis, 
and decays to the ground state $\ket{g}$. The Hamiltonian of the system is
\begin{multline}
H = \sum_k \hbar \omega_k \hat{a}_k^{\dag} \hat{a}_k + \sum_{j=1}^N \sum_{\mu =g,s,e} \hbar \omega_{\mu} \ket{\mu}_j\bra{\mu}  \\ 
- \sum_{j=1}^N \Bigl[\hbar \Omega e^{i (\bm{k}_c \cdot \bm{r}_j - \omega_c t)} \ket{e}_j\bra{s} \\
+  \bm{\wp}_{eg} \cdot \hat{E}(\bm{r}_j)  \ket{e}_j\bra{g} + \mathrm{H. c}\Bigr] , 
\end{multline}
where the first term on the r.h.s. is the Hamiltonian for the field modes with energies $\hbar \omega_k$, 
the second term corresponds to the Bohr energies $\hbar \omega_{\mu}$ of the atomic levels $\ket{\mu}$ ($\mu = g,s,e$),
the third term describes the interaction of the atoms with the coupling laser with frequency $\omega_c$ and wavevector
$\bm{k}_c \parallel \hat{\bm{z}}$, and the last term describes the coupling of the atoms to the free-space radiation field
with the dipole moment $\bm{\wp}_{eg}$ on the transition $\ket{e} \to \ket{g}$. We set the energy of the ground state to zero, 
$\hbar \omega_g=0$, and assume that $\omega_s \ll \omega_e$ ($\omega_c \simeq \omega_e$). 

The state vector of the system with the single atomic or photonic excitation can be expanded as
\begin{eqnarray*}
\ket{\Psi} &=& \sum_j c_j e^{-i \omega_s t} \ket{s_j} \otimes \ket{0} + \sum_j b_j e^{-i \omega_e t} \ket{e_j}\otimes \ket{0} 
\nonumber \\ & &
+ \ket{G} \otimes \sum_k a_k e^{-i \omega_k t} \ket{1_k} , 
\end{eqnarray*}
where  $\ket{s_j} \equiv \ket{g_1,g_2,\ldots,s_j, \ldots, g_N}$,
$\ket{e_j} \equiv \ket{g_1,g_2,\ldots,e_j, \ldots, g_N}$, 
$\ket{G} \equiv \ket{g_1,g_2,\ldots,g_N}$, and 
$\ket{1_k} \equiv \hat{a}_k^{\dag}\ket{0}$ is the state of the radiation field with a single photon in mode $k$.
The state vector is normalized as $\braket{\Psi}{\Psi} = \sum_{j} (|c_j|^2 + |b_j|^2) + \sum_k |a_k|^2 =1$,
and it evolves according to the Schr\"odinger equation $\partial_t \ket{\Psi} = -\frac{i}{\hbar} H \ket{\Psi}$, 
leading to a set of equations for the atomic amplitudes  
\begin{subequations}
\label{eqs:cjbj}
\begin{eqnarray}
\partial_t c_j &=& i \Omega^* e^{-i\bm{k}_c \cdot \bm{r}_j } b_j e^{i \Delta_c t}, \label{eqs:cj} \\
\partial_t b_j &=& i \Omega  e^{i \bm{k}_c \cdot \bm{r}_j} c_j e^{- i \Delta_c t}  + i \sum_k g_k(\bm{r}_j) a_k e^{i(\omega_e-\omega_k)t} , \qquad
\label{eqs:bj}
\end{eqnarray}
\end{subequations}
with $\Delta_c = \omega_c - \omega_e$, and an equation for the field amplitudes cast in the integral form 
\begin{equation}
a_k(t) = i \sum_j g_k^*(\bm{r}_j) \int_0^t \!\! dt' b_j(t') e^{i(\omega_k-\omega_e)t'} , \label{eq:ak}
\end{equation} 
where $g_k(\bm{r}_j) = \frac{\bm{\wp}_{eg} \cdot u_k(\bm{r}_j)}{\hbar}$. 

\subsection{Atoms}

We substitute Eq.~(\ref{eq:ak}) into Eq.~(\ref{eqs:bj}) and use the basis of the plane waves 
$u_k(\bm{r}) = \hat{\bm{\veps}}_{\bm{k},\sigma} \sqrt{\frac{\hbar \omega_k}{2 \eps_0 V}} e^{i \bm{k} \cdot \bm{r}}$ 
within the quantization volume $V$. We then replace the summation over the modes by an integration, and 
use the Born-Markov approximation to eliminate the radiation field \cite{ScullyZubary1997,PLDP2007}, 
while neglecting the field-mediated interatomic interactions assuming sufficiently large mean interatomic 
distance $k \expv{|\bm{r}_j-\bm{r}_i|} \gg 1$ \cite{Lehmberg1970,Thirunamachandran,Miroshnychenko2013}.
We thus obtain the usual spontaneous decay rate 
$\Gamma = \frac{1}{4 \pi \eps_0} \frac{4 k_e^3 |\wp_{eg}|^2}{3 \hbar}$ and the Lamb shift 
of level $\ket{e}$ that can be incorporated into $\omega_e$. Equations~(\ref{eqs:cjbj}) reduce to 
\begin{subequations}
\label{eqs:cjbjG}
\begin{eqnarray}
\partial_t c_j &=& i \Omega^* e^{-i\bm{k}_c \cdot \bm{r}_j } b_j e^{i \Delta_c t}, \label{eqs:cjG} \\
\partial_t b_j &=& - \Gamma/2 \, b_j + i \Omega e^{i \bm{k}_c \cdot \bm{r}_j} c_j e^{- i \Delta_c t}. \label{eqs:bjG}
\end{eqnarray}
\end{subequations}

We next assume a resonant laser $\Delta_c = 0$ with sufficiently weak Rabi frequency $|\Omega| \ll \Gamma$ and
set $\partial_t b_j = 0$, obtaining $b_j \simeq i \frac{\Omega}{\Gamma/2}  e^{i \bm{k}_c \cdot \bm{r}_j}c_j$. Substituting
this into Eq.~(\ref{eqs:cjG}) and performing the integration, we finally obtain 
\begin{gather}
b_j(t)  \simeq  i c_j(0) \beta(t) e^{i \bm{k}_c \cdot \bm{r}_j} , \label{eq:bjt} \\
\beta(t)  \equiv  \frac{\Omega(t)}{\Gamma/2} \exp \left[ - \int_0^t \!\! dt' \frac{|\Omega(t')|^2}{\Gamma/2} \right], \nonumber
\end{gather}
with the initial condition $b_j(0) = 0$ and $c_j(0) = \frac{e^{i\phi_j}}{\sqrt{N}} \; \forall \; j \in \{ 1,N \}$,
where we included a spatial phase $\phi_j$ of the single-excitation spin wave. 
In Sec.~\ref{sec:results} we will analyze the influence of this phase on the photon collection into Gaussian optical modes.

\subsection{Field}

We may define the wavefunction of the emitted single-photon field via 
$E(\bm{r},t) \equiv \bra{0} \bra{G} \hat{E}(\bm{r}) \ket{\Psi (t)}$ \cite{ScullyZubary1997,PLDP2007}. 
Assuming an isotropic dipole moment, in Appendix \ref{app:sphwf} we show that 
$E(\bm{r},t) \propto \sum_j \frac{e^{-i \om_e (t-|\bm{r} -\bm{r}_j|/c)}}{|\bm{r} -\bm{r}_j|} b_j(t - |\bm{r} -\bm{r}_j|/c)$, 
and for an atomic ensemble with density $\rho(\bm{r})$, the emitted field is
\begin{multline}
E(\bm{r},t) = \frac{\wp_{eg} k_e^2}{\sqrt{3} \, 2\pi \eps_0} 
\int \! d^3 r' \rho(\bm{r}') \, \frac{e^{-i \om_e (t-|\bm{r} -\bm{r}'|/c)}}{|\bm{r} -\bm{r}'|} \\ 
\times b_{\bm{r}'}(t - |\bm{r} -\bm{r}'|/c) \,. \label{eq:Ertex}
\end{multline}

\subsubsection*{Paraxial optics}

We assume that $N \gg 1$ atoms in a harmonic trap have a cylindrically symmetric, normal density distribution 
\begin{equation} 
\rho(\bm{r}) = N \frac{e^{-(x^2+ y^2)/2\sigma_{\perp}^2 - z^2/2\sigma_z^2}}{(2 \pi)^{3/2} \sigma_{\perp}^2 \sigma_z}   \label{eq:rhor}
\end{equation}
with the width $\sigma_{\perp}$ and length $\sigma_z$ (standard deviations).
Consistent with the assumption of closely spaced energy levels $\ket{g}$ and $\ket{s}$ ($\omega_c \simeq \omega_e$),
we have $k_c \simeq k_e$ such that $|k_c - k_e| \ll 1/\sigma_{z}$.
The radiation emitted in the phase-matched direction $\bm{k}_e \simeq \bm{k}_c \parallel \hat{\bm{z}}$ 
is collected by paraxial optics which feeds it into a waveguide with a single transverse mode and a continuum of
1D modes $k$, see Fig.~\ref{fig:Scheme}(b). The atomic ensemble is placed at the focus of the collecting optics
with the beam waist $w_0$. The corresponding field modes that couple to the waveguide are then the Gaussian modes 
\begin{subequations}
\label{eq:Gfk}
\begin{gather}
v_k(\bm{r}) = \sqrt{\frac{\hbar \om_k}{2 \eps_0 AL}} \, \upsilon_{k}(\bm{r}), \\  
\upsilon_{k}(\bm{r}) \equiv \frac{\zeta_k}{q_k^*(z)} \exp \left[ i k \left( z + \frac{x^2+y^2}{2q_k^*(z)} \right) \right] \\
= i \frac{w_0}{w(z)} \exp \left[ i k z - \frac{x^2+y^2}{w(z)^2} + i \frac{k(x^2+y^2)}{2R(z)} - i \phi_{\mr{Gouy}} \right] , \nonumber 
\end{gather}
\end{subequations}
where $A = \pi w_0^2/2$ is the cross-section at the focus $z=0$ and $L$ is the quantization length, 
$\zeta_k = k w_0^2/2$ is the Rayleigh length and $q_k(z) = z + i \zeta_k$ is the complex beam parameter,
$w(z) = w_0\sqrt{1+z^2/\zeta_k^2} $ is the transverse beam size at position $z$, 
$R(z) = z+\zeta_k^2/z $ is the radius of curvature of the phase front, and 
$\phi_{\mr{Gouy}} = \arctan(z/\zeta_{k_e})$ is the Gouy phase.
 
Our aim is to maximize the collection of radiation emitted by the atoms by the paraxial optical elements. 
If we, however, attempt to calculate the overlap between the single-photon field of Eq.~(\ref{eq:Ertex}) 
and the resonant $k=k_e$ Gaussian mode of Eq.~(\ref{eq:Gfk}) as a volume integral 
$\left| \int \! d^3 r \, v_{k_e}^*(\bm{r}) E(\bm{r}) \right|^2$, it will diverge inside 
the atomic ensemble, where $\rho(\bm{r}') \neq 0$, due to the $|\bm{r} -\bm{r}'|^{-1}$ term.

\subsubsection*{Photon number in the forward Gaussian modes}

An alternative, more tractable approach that we use here is to calculate directly 
the total number of photons $n(t)$ emitted into the 1D continuum $\{ k \}$ of 
the forward Gaussian modes $v_k(\bm{r})$ all having the same waist $w_0$ at $z=0$. 
We have 
\begin{gather}
n(t) = \sum_k |a_k(t)|^2, \label{eq:nt} \\ 
a_k(t) = i \sum_j g^*_k(\bm{r}_j) \int_0^t \!\! dt' \, b_j(t') e^{i (\omega_k - \omega_e) t'} ,  \nonumber
\end{gather}
where $g_k(\bm{r}_j) =  \frac{\wp_{eg}}{\hbar} v_k(\bm{r}_j)$ is now the coupling strength of atom $j$ 
at position $\bm{r}_j$ to the $k$'s Gaussian mode of Eq.~(\ref{eq:Gfk}).

In Appendix \ref{app:satphot} we show that for a single atom placed at the origin, $\bm{r}_j =0$,
the amount of radiation collected by the Gaussian paraxial optics,
\begin{equation}
n(t) = \frac{\varsigma}{2A} \Gamma \int_0^t dt' \, |b(t')|^2 ,
\end{equation}
is proportional to the ratio of the atomic absorption cross-section $\varsigma = 3 \lambda_e^2/2\pi$
to the cross-section $A = \pi w_0^2/2$ of the focused Gaussian beam at the atomic position. 
For many atoms $N$ with the density distribution of Eq.~(\ref{eq:rhor}), we have 
\begin{multline}
n(t) =  \frac{|\wp_{eg}|^2}{2 \pi \hbar \eps_0 c A} \int_0^{\infty} \!\!\! d \omega_k \, \omega_k 
\sum_{j,j'}^N \upsilon_k^*(\bm{r}_j) \upsilon_k(\bm{r}_{j'}) 
 \\ \times
\iint_0^t \!\! dt' dt''  \, b_j(t') b_{j'}^*(t'')  e^{i (\omega_k - \omega_e) (t'-t'')} 
 \\ 
=  \frac{|\wp_{eg}|^2}{2 \pi \hbar \eps_0 c A} \int_0^{\infty} \!\!\! d \omega_k \, \omega_k 
\iint \! d^3 r \, d^3 r' \rho(\bm{r}) \upsilon_k^*(\bm{r}) \rho(\bm{r}') \upsilon_k(\bm{r}') 
 \\ \times
\iint_0^t \!\!dt' dt''  \, b_{\bm{r}}(t') b_{\bm{r}'}^*(t'')  e^{i (\omega_k - \omega_e) (t'-t'')}.
\end{multline}
Substituting here Eq.~(\ref{eq:bjt}) with $c_{\bm{r}}(0) = e^{i\phi(\bm{r})} /\sqrt{N}$, where $\phi(\bm{r})$
is a spatial phase profile of the stored spin-wave, we have 
\begin{eqnarray}
n(t) &=&  \frac{|\wp_{eg}|^2}{2 \pi \hbar \eps_0 c A} \frac{1}{N} \iint_0^t \! dt' dt'' \, \beta(t') \beta(t'') 
\nonumber \\ & & \qquad \times
\int_0^{\infty} \!\!\! d \omega_k \, \omega_k \, \Xi_k  e^{i (\omega_k - \omega_e) (t'-t'')} , \\
& & \quad \Xi_k \equiv  \left|\int \! d^3 r \rho(\bm{r}) \upsilon_k^*(\bm{r}) e^{i \bm{k}_c \cdot \bm{r} + i\phi(\bm{r})} \right|^2 . \nonumber 
\end{eqnarray}
We can proceed along the Weisskopf-Wigner approximation \cite{ScullyZubary1997,Miroshnychenko2013}
by replacing $\omega_k$ and $\Xi_k$ by their resonant values $\omega_e$ and $\Xi_{k_e}$ and pulling them out 
of the frequency integral, which, upon extending the lower limit of integration, reduces to $2\pi \delta(t'-t'')$. 
We then obtain   
\begin{equation}
n(t) =  \frac{\varsigma}{2A}|\xi|^2 \, N \Gamma B(t) , \label{eq:ntens}
\end{equation}
where $B(t) \equiv  \int_0^t dt' \beta^2(t')$ contains the time-dependence of the photon envelope, 
while the overlap integral is
\begin{multline}
\xi \equiv \frac{1}{(2 \pi)^{3/2} \sigma_{\perp}^2 \sigma_z}
\int \! d^3 r \exp \left(i \phi(\bm{r}) -\frac{x^2+ y^2}{2\sigma_{\perp}^2} - \frac{z^2}{2\sigma_z^2} \right)
 \\  \times
\frac{\zeta_{k_e}}{q_{k_e}(z)} \exp \left(- i k_e \frac{x^2+y^2}{2q_{k_e}(z)} \right) ,  \label{eq:xi}
\end{multline}
where we used that $k_e \simeq k_c$ ($\bm{k}_c \parallel \hat{\bm{z}}$). 
Note that in Eq.~(\ref{eq:ntens}) the geometric factor is
\begin{equation}
\frac{\varsigma}{2A}|\xi|^2 = \frac{6 |\xi|^2}{(k_e w_0)^2} \equiv  \mc{G} \, . 
\end{equation}

Equations (\ref{eq:ntens})-(\ref{eq:xi}) are the central equations of this paper. 
Our aim is to find the optimal value for the Gaussian beam waist $w_0$ that maximizes 
the probability of collecting the photon $n(\infty) \propto \mc{G} N$ emitted by 
an ensemble of $N$ trapped atoms with the given width and length $\sigma_{\perp},\sigma_z$ 
(standard deviations) of the density distribution. 

\section{Optimal photon collection}
\label{sec:results}

We now present the results of our calculations for three different cases 
of the initial spatial phase distribution in the atomic cloud centered at the beam focus. 

\subsubsection{Uniform spatial phase}

Assume that the initial atomic spin-wave has a uniform spatial phase 
\begin{equation}
\phi(\bm{r}) = 0 \, \forall \, \bm{r} . \label{eq:phi0}
\end{equation}
Consider first the case of the atom cloud with the spatial dimensions 
much smaller than the Rayleigh length, $\sigma_\perp, \sigma_z \ll \zeta_{k_e}$. 
The mode function appearing in Eg.~(\ref{eq:xi}) can then be approximated as
\[
\frac{\zeta_{k_e}}{q_{k_e}(z)} \exp \left( -i k_e  \frac{x^2+y^2}{2q_{k_e}(z)} \right) 
\simeq \exp \left( -\frac{x^2+y^2}{w_0^2} + i \frac{z}{\zeta_{k_e}}  \right) ,
\]
where we have used that near the focus the beam waist is $w_0$, the beam radius of curvature $R$ tends to infinity, 
while the Gouy phase can be linearized as $\phi_{\mr{Gouy}} \simeq z/\zeta_{k_e}$. 
Equation~(\ref{eq:xi}) reduces to a Gaussian integral and we obtain 
\begin{eqnarray}
|\xi|^2 & \simeq & \frac{w_0^4}{(2\sigma_\perp^2 + w_0^2)^2} \exp \left[ -\left( \frac{2 \sigma_z}{k_e w_0^2} \right)^2 \right]  
\nonumber \\ 
&=& \frac{\bar w_0^4}{(2\bar \sigma_\perp^2+\bar w_0^2)^2} \exp \left[ -\left( \frac{2\bar \sigma_z}{\bar w_0^2} \right)^2 \right]  ,
\label{eq:xiSmallcl}
\end{eqnarray}
where the dimensionless quantities are defined as $\bar{\sigma}_{z,\perp} \equiv k_e \sigma_{z,\perp}$ and $ \bar{w}_0 \equiv k_e w_0$. 
The optimal beam waist that maximizes the geometric factor $\mc{G} = 6 |\xi|^2/\bar w_0^2$ is 
\begin{equation}
\bar{w}^{\text{max}}_0 = \sqrt{\frac23 \left[\bar \sigma_\perp^2 + P(\bar \sigma_\perp, \bar \sigma_z) + 
\frac{\bar \sigma_\perp^4+6\bar \sigma_z^2}{P(\bar \sigma_\perp, \bar \sigma_z)}\right]} , \label{eq:WaistOpt} 
\end{equation}
\begin{multline*}
P(\bar \sigma_\perp, \bar \sigma_z) =  \Bigl[\bar{\sigma}_\perp^6 + 36 \bar \sigma^2_\perp \bar \sigma_z^2
\\ + 3\sqrt{6\bar \sigma_z^2\left(\bar \sigma_\perp^8+22\bar \sigma^4_\perp \bar \sigma_z^2-4\bar \sigma_z^4\right)}\Bigr]^{1/3} . 
\end{multline*}
In the limit of $ \sigma_z \to 0$, the photon collection efficiency is maximized for $w^{\text{max}}_0 = \sqrt{2} \sigma_\perp$, 
with $|\xi|^2 \to 1/4$ and thus $\mc{G} \simeq 3/4 \bar{\sigma}_\perp^2$. 
This is of course an intuitive result meaning that the waist of the Gaussian beam should be equal 
to the width of the atomic ensemble. As will be seen below, this result also applies to the atomic ensembles
with sufficiently large width, which is, however, not optimal for photon collection by the paraxial Gaussian optics. 

\begin{figure*}[t]
\includegraphics[width=1.0\linewidth]{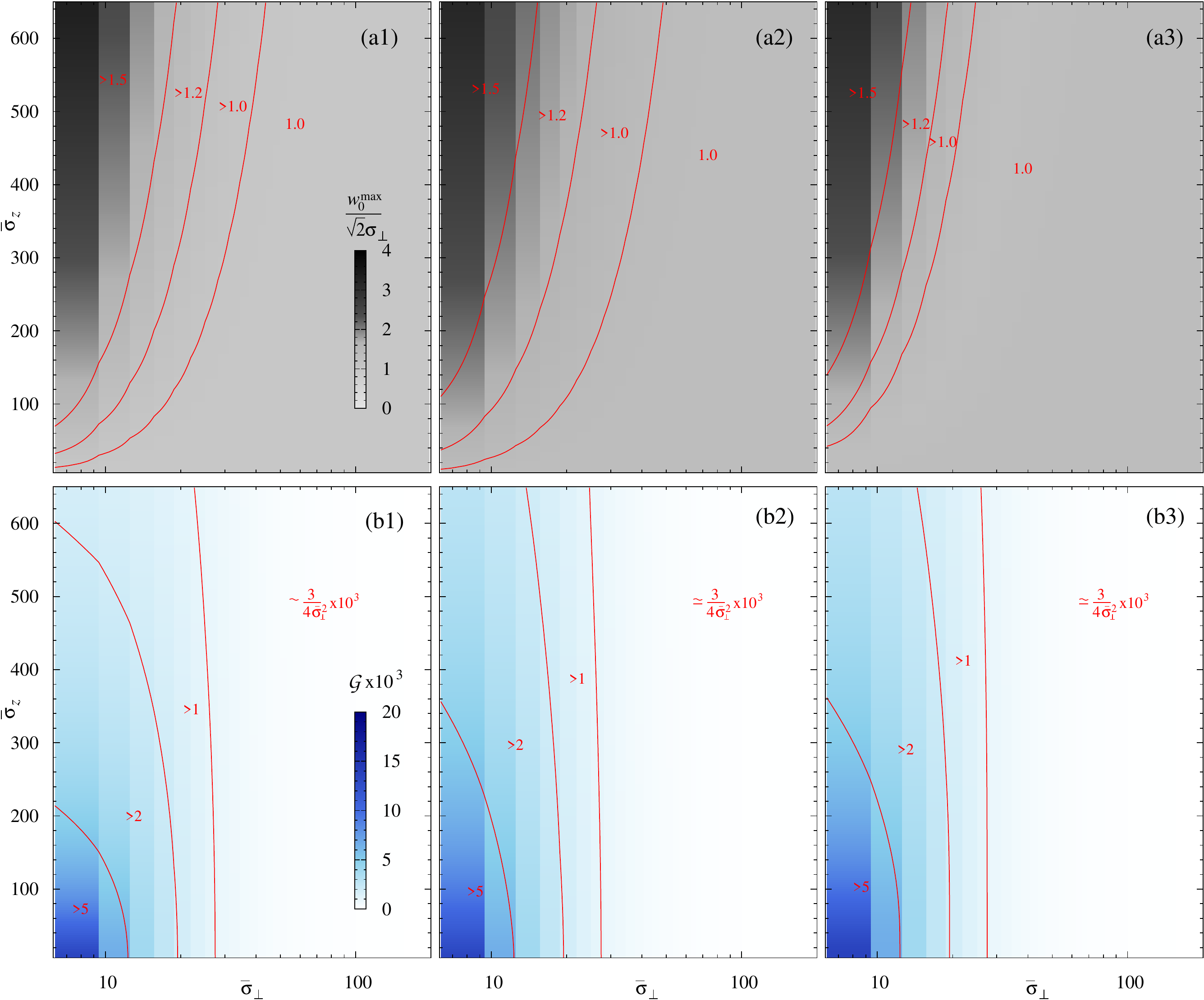}
\caption{Optimal waist of the Gaussian beam $\bar{w}^{\text{max}}_0$ divided by the width of the atomic ensemble 
$\sqrt{2} \sigma_{\perp}$ (upper panels a) and 
the corresponding maxima of the geometric factor $\mc{G} \equiv 6|\xi|^2/(\bar{w}^{\text{max}}_0)^2$ for $N=10^3$
atoms (lower panels b) 
versus the dimensionless width and length $\bar{\sigma}_{\perp,z} = k_e \sigma_{\perp,z}$ of the atomic cloud,
for the cases of the spatial phase $\phi(\bm{r})$ of the atomic spin wave given by: 
Eq.~(\ref{eq:phi0}), uniform phase, (a1), (b1);
Eq.~(\ref{eq:phiGouy}), Gouy phase compensation, (a2), (b2); and 
Eq.~(\ref{eq:phiGauss}),  full spatial phase compensation, (a3), (b3).
For the values of $\bar{\sigma}_{\perp,z}$ where $\bar{w}^{\text{max}}_0 \simeq \sqrt{2}\bar{\sigma}_{\perp}$ 
we also have $|\xi|^2 \simeq 1/4$ and thus $\mc{G} \simeq 3/4 \bar{\sigma}_{\perp}^2$. }
\label{fig:w0n}
\end{figure*}

When the atomic cloud is larger than the Rayleigh length, the integral of Eq.~(\ref{eq:xi}) can be evaluated as
\begin{eqnarray}
\xi^* &=& \frac{\zeta_{k_e}}{(2\pi)^{3/2} \,\sigma_\perp^2 \sigma_z} \int \! dz \frac{e^{-\frac{z^2}{2\sigma_z^2}}}{q^*_{k_e}(z)} 
\nonumber \\ & & \qquad \times
\int \! dx dy \exp\left[ -(x^2+y^2) \left(\frac{1}{2\sigma_\perp^2} -i \frac{k_e}{2q^*_{k_e}(z)}\right) \right] 
\nonumber \\
&=& \frac{\zeta_{k_e}}{(2\pi)^{1/2}\,\sigma_z} \int dz \frac{e^{-\frac{z^2}{2\sigma_z^2}}}{z-i\zeta_{k_e}- i k_e \sigma_\perp^2} 
\nonumber \\
&=& i\sqrt{\frac{\pi}{8}} \frac{k_e w_0^2}{\sigma_z} \,e^{\frac{\left(k_e w_0^2/2+k_e \sigma_\perp^2 \right)^2}{2\sigma_z^2}}
\mathrm{erfc} \left(\frac{k_e w_0^2/2+ k_e \sigma_\perp^2}{\sqrt{2} \sigma_z}\right)\nonumber \\
&=& i\sqrt{\frac{\pi}{8}}\frac{\bar w_0^2}{\bar\sigma_z} \,e^{\frac{\left(\bar w_0^2/2+\bar\sigma_\perp^2 \right)^2}{2\bar\sigma_z^2}}
\mathrm{erfc} \left(\frac{\bar w_0^2/2+\bar\sigma_\perp^2}{\sqrt{2}\bar\sigma_z}\right) . \label{eq:xi2max}
\end{eqnarray}
In Fig.~\ref{fig:w0n}(a1), we show the optimal values of $\bar{w}^{\text{max}}_0$ that maximize 
$\mc{G}$ for various values of $\bar{\sigma}_z, \bar{\sigma}_{\perp}$. 
The corresponding maxima for $\mc{G}$ are shown in Fig.~\ref{fig:w0n}(b1). 
We notice that for a wide range of values of $\bar{\sigma}_z, \bar{\sigma}_{\perp}$ 
the optimal waist of the Gaussian beam is again $\bar{w}^{\text{max}}_0 \simeq \sqrt{2} \bar{\sigma}_{\perp}$
and the corresponding maximum for the geometric factor is $|\xi|^2 \simeq 1/4$ and 
thus $\mc{G} = 6 |\xi|^2/(\bar{w}^{\text{max}}_0)^2 \simeq 3/4 \bar{\sigma}_\perp^2$. 
Only for sufficiently narrow and not too long ensembles, $\bar{\sigma}_{\perp} \lesssim 10$ ($\sigma_{\perp} \sim \lambda_e$) 
and $\bar{\sigma}_z \lesssim 200$  ($\sigma_{z} \lesssim 30 \lambda_e$), we obtain a significantly
better photon collection efficiency $n(\infty) \propto \mc{G} N \gtrsim 5$ (assuming $N=10^3$ atoms)
into the appropriately focused ($w_0 = w^{\text{max}}_0$) Gaussian beams.

%%%%%%%%%%%%%%%%%%%%%%%%%%%%%%%%%%%%%%%%
%%%%%%%%%%%%%%%%%%%%%%%%%%%%%%%%%%%%%%%%

\subsubsection{Spatial phase compensating the Gouy phase}

Recall that the Gouy phase $\phi_{\mr{Gouy}}(z) = \arctan(z/\zeta_{k_e})$ of the Gaussian beam varies with $z$. 
We may therefore further improve the photon collection efficiency if we imprint onto the stored atomic spin-wave 
a spatial phase that would compensate the Gouy phase, 
\begin{equation}
\phi(\bm{r}) = -\phi_{\mr{Gouy}}(z) . \label{eq:phiGouy}
\end{equation}
This can be achieved, e.g., by using spatially varying electric or magnetic fields or ac Stark shifts 
induced by off-resonant lasers \cite{Korzeczek2020}. Now the integral of Eq.~(\ref{eq:xi}) can be evaluated as 
\begin{eqnarray}
\xi^* &=& \frac{\zeta_{k_e}}{(2\pi)^{3/2} \,\sigma_\perp^2 \sigma_z} \int \! dz \frac{e^{-\frac{z^2}{2\sigma_z^2}}}{|q^*_{k_e}(z)|} 
\nonumber \\ & & \qquad \times
\int \! dx dy \exp\left[ -(x^2+y^2) \left(\frac{1}{2\sigma_\perp^2} -i \frac{k_e}{2q^*_{k_e}(z)}\right) \right] 
\nonumber \\
&=& \frac{\zeta_{k_e}}{(2\pi)^{1/2}\,\sigma_z} \int dz \frac{e^{-\frac{z^2}{2\sigma_z^2}}}{|q^*_{k_e}(z)|} \cdot \frac{q^*_{k_e}(z)}{q^*_{k_e}(z)-ik_e \sigma_\perp^2} \nonumber \\ 
%&=& \frac{\zeta_{k_e}}{(2\pi)^{1/2}\,\sigma_z} \int dz \frac{e^{-\frac{z^2}{2\sigma_z^2}}}{|q^*_{k_e}(z)|} \cdot \frac{|q^*_{k_e}(z)| e^{- i \phi_{\mr{Gouy}}(z)}}{q^*_{k_e}(z)-ik_e \sigma_\perp^2} \nonumber \\ 
&=& \frac{\zeta_{k_e}}{(2\pi)^{1/2}\,\sigma_z} \int dz \frac{e^{-\frac{z^2}{2\sigma_z^2} - i \phi_{\mr{Gouy}}(z)}}{z-i\zeta_{k_e}- i k_e \sigma_\perp^2} . \label{eq:xiComp}
\end{eqnarray}
Alternatively, we can write
\begin{eqnarray}
\!\! \xi^* &=& \frac{i}{(2\pi)^{3/2}\,\sigma_\perp^2 \sigma_z} \int \! dz \frac{w_0}{w(z)} e^{-\frac{z^2}{2\sigma_z^2}} \nonumber \\
& & \; \times \int \! dx dy \exp \left[ -(x^2+y^2)\left( \frac{1}{w(z)^2} + \frac{1}{2 \sigma_\perp^2} - i \frac{k_e}{2R(z)} \right) \right] \nonumber \\
&=&\frac{i w_0}{\sqrt{2\pi} \sigma_z}\int \! dz \frac{e^{-\frac{z^2}{2\sigma_z^2}} w(z)R(z)}{2R(z)\sigma_\perp^2 + w(z)^2 R(z)-i k_e w(z)^2 \sigma_\perp^2 } .
\end{eqnarray}
We numerically maximize $|\xi|^2/(\bar{w}_0)^2$ for different values of $\bar{\sigma}_z, \bar{\sigma}_{\perp}$, and 
show the resulting $\bar{w}^{\text{max}}_0$ in Fig.~\ref{fig:w0n}(a2) and the corresponding maxima for $\mc{G}$ in Fig.~\ref{fig:w0n}(b2).
Note that while the overall behavior of the geometric factor is similar to the previous case of uniform spatial phase, 
we nevertheless obtain good photon collection efficiency $n(\infty) \propto \mc{G} N \gtrsim 5$ ($N=10^3$) 
even for longer atomic clouds $\bar{\sigma}_z \lesssim 300$  ($\sigma_{z} \lesssim 50 \lambda_e$) since we compensate 
for the Gouy phase that changes the sign across the focal point $z=0$. In other words, when a photon emitted 
from a highly elongated atomic ensemble, $\sigma_z \gg \zeta_k$ ($\bar{\sigma}_z \gg \bar{w}_0^2/2$), with  
a uniform spatial phase $\phi(\bm{r}) = 0$ is collected into a focused Gaussian mode, 
the photon amplitudes originating from the regions of $z<0$ and $z>0$ interfere partially destructively, 
while for the spatial phase $\phi(\bm{r}) = -\phi_{\mathrm{Gouy}}(z)$ this interference is constructive.  

\subsubsection{Full spatial phase compensation of a Gaussian beam}

We finally examine the case of the spatial phase of the atomic spin wave having the same spatial dependence as that of a 
focused Gaussian beam,
\begin{equation}
\phi(\bm{r}) = k \frac{x^2 +y^2}{2R(z)} - \phi_{\mr{Gouy}}(z) , \label{eq:phiGauss}
\end{equation}
which would correspond to a photon from a Gaussian mode stored in an atomic ensemble, 
e.g., via stopping a dark-state polariton in the EIT regime \cite{EITrev2005}.
For the integral of Eq.~(\ref{eq:xi}) we then obtain  
\begin{eqnarray}
\xi^* &=&  \frac{i}{(2\pi)^{3/2}\,\sigma_\perp^2 \sigma_z} \int \! d^3 r 
\exp \left(-\frac{x^2+ y^2}{2\sigma_{\perp}^2} - \frac{z^2}{2\sigma_z^2} \right)  
\nonumber \\ & & \quad \times 
\frac{w_0}{w(z)} \exp \left( -\frac{x^2+y^2}{w^2}\right) 
\nonumber \\
&=& i \frac{w_0}{\sqrt{2\pi} \sigma_z}\int \! dz e^{-\frac{z^2}{2\sigma_z^2}}\frac{w(z)}{w(z)^2+2\sigma_{\perp}^2} .
\end{eqnarray}
For a small atomic cloud, $w(z) \simeq w_0 $, we have the analytic result $|\xi|^2 = w_0^4/(w_0^2+2\sigma_\perp^2)^2$, 
which is the same as in Eq.~(\ref{eq:xiSmallcl}) in the limit of $ \sigma_z \rightarrow 0 $. For larger cloud sizes,
we again maximize numerically $|\xi|^2/\bar w_0^2$ for different values of $\bar{\sigma}_z, \bar{\sigma}_{\perp}$, 
obtaining $\bar{w}^{\text{max}}_0$ and the corresponding maxima for $\mc{G}$ which are shown, respectively, 
in Fig.~\ref{fig:w0n}(a3) and (b3). Even though $\bar{w}^{\text{max}}_0 \simeq \sqrt{2} \bar{\sigma}_{\perp}$
for even smaller values of $\bar{\sigma}_{\perp}$ as compared to case 2 above, the maximum for the photon collective 
efficiency $n(\infty) \propto \mc{G} N$ has nearly the same dependence on $\bar{\sigma}_z, \bar{\sigma}_{\perp}$
as that with only the Gouy phase compensation. The physical reason for this result is that the curvature of the phase front 
of a Gaussian mode plays only a minor role in narrow atomic ensembles yielding the largest photon collection efficiency.

\section{Discussion and Conclusions}
\label{sec:conclud}

Cold, trapped atoms can strongly interact with radiation of different frequencies and therefore  
be used for coherently interfacing microwave and optical systems and components. 
When converting atomic excitations to photons, an important consideration for efficient 
collection of radiation is the directionality of the emitted photons. 
A well-known approach for efficient photon collection is to place the atoms or other emitters into a resonant cavity 
with sufficiently strong coupling so that the Purcell enhancement of the photoemission rate into the cavity mode 
largely exceeds the free-space spontaneous decay rate with the photon emitted in a random direction \cite{PLDP2007}. 
In contrast, using wave-mixing or stimulated Raman processes, a coherent spin wave of a large atomic ensemble 
can be converted into a photon emitted predominantly into the phase-matched direction with well-defined 
temporal and spatial profile \cite{Petrosyan2018,Petrosyan2019}.
 
In this paper, we have addressed the question of how to optimally collect this photon 
created from the atomic spin wave by a stimulated Raman process. 
Assuming an atomic ensemble with Gaussian density distribution -- typical for cold atoms in harmonic traps --
we have found that the emitted photon is best mode-matched by a Gaussian optical mode with the waist 
equal to the width of the atomic ensemble, provided it is not too small (larger than the wavelength) 
[see Fig.~\ref{fig:w0n}(a)].  
For narrower ensembles, the optimal waist of the Gaussian beam should be larger, but the photon 
collection efficiently decreases for highly elongated ensembles [see Fig.~\ref{fig:w0n}(b)]. 
The main reason for this is the reduction of the spatial overlap between the atomic cloud and 
the tightly focused Gaussian beam whose transverse width rapidly increases away from the focus. 
Hence, compact, i.e. narrow and short, atomic ensembles offer the best mode matching with appropriately 
focused Gaussian beams. But since increased probability of directional photoemission also requires 
a large number of atoms $N$ and large optical depth of the ensemble [see Eq.~(\ref{eq:ntens})], 
compact atomic ensembles will have high density, which necessitates considerations of atom-atom 
interactions and multiple scattering \cite{Petrosyan2021}.

%\section*{Acknowledgments}
%The work of A.K., P.D. and A.V. was supported by the National Research, Development and Innovation Office of Hungary 
%(Project Nos. K115624, 2017-1.2.1-NKP-2017-00001). 
%A.V. was also supported by the J\' anos Bolyai Research Scholarship of the Hungarian Academy of Sciences. 
%T.B. and N.T.Z. acknowledge support from the Carlsberg Foundation and the Independent Research Fund Denmark.
%J.F. and D.P. were supported by the DFG SPP 1929 GiRyd and DFG Project No. 394243350.
%D.P. was in addition supported by the EU QuanERA Project PACE-IN and by 
%the Alexander von Humboldt Foundation in the framework of the Research Group Linkage Programme.

\appendix 

\section{Single-photon wavefunction}
\label{app:sphwf}

The wavefunction of the emitted single-photon field can be defined via \cite{ScullyZubary1997,PLDP2007,Miroshnychenko2013}
\begin{eqnarray}
E(\bm{r},t) & \equiv & \bra{0} \bra{G} \hat{E}(\bm{r}) \ket{\Psi (t)} = \sum_k u_k (\bm{r}) a_k(t) e^{-i \omega_k t}  
\nonumber \\ 
&=& i \sum_k u_k (\bm{r}) \sum_j g_k^*(\bm{r}_j) \int_0^t \!\! dt' b_j(t') e^{-i \omega_k (t-t') - i \omega_e t'} . \qquad
\end{eqnarray}
For simplicity, we assume an isotropic dipole moment 
$\bm{\wp}_{eg} \cdot \hat{\bm{\varepsilon}} = \wp_{eg}/\sqrt{3} \, \forall \, |\hat{\bm{\varepsilon}}| = 1$ 
and use the basis of the plane waves, leading to  
\begin{eqnarray}
E(\bm{r},t) &=& i \frac{\wp_{eg}}{\sqrt{3} (2\pi)^3 \eps_0} \sum_j  \int_0^t \!\! dt' b_j(t') e^{- i \omega_e t'}
\nonumber \\ & & \times
\int_0^{\infty} \!\! dk k^2 \omega_k  e^{-i \omega_k (t-t')} \int_{4\pi} \!\! d \Omega_k e^{i \bm{k} \cdot (\bm{r} -\bm{r}_j) } , \qquad
\label{eq:Ertplw}
\end{eqnarray}
where we replaced the summation over the modes by an integration,  
$\sum_k \to 2 \frac{V}{(2\pi)^3} \int \! d^3 k = 2 \frac{V}{(2\pi)^3}  \int_0^{\infty} \! d k \, k^2 \int_{4\pi} \! d \Omega_k$
with the factor of 2 accounting for the two orthogonal photon polarizations $\sigma=1,2$ for each $\bm{k}$
($\hat{\bm{\veps}}_{\bm{k},\sigma} \perp \bm{k}$). The integration over 
the $4\pi$ solid angle with $d \Omega_k = \sin \theta d \theta d \varphi$ leads to 
$4\pi \frac{\sin (k |\bm{r} -\bm{r}_j|) }{k |\bm{r} -\bm{r}_j|} 
= -i \frac{2\pi c}{\omega_k|\bm{r} -\bm{r}_j|} (e^{i k |\bm{r} -\bm{r}_j|} - \mathrm{c.c})$.
We substitute this into the above equation, assume that during the photon emission $k$ 
is peaked around the atomic resonance $k_e = \omega_e/c$ and pull $k_e^2$ out of the integral, 
and extend the lower limit of integration over $k$ to $-\infty$, as in the Weisskopf-Wigner 
approximation \cite{ScullyZubary1997,Miroshnychenko2013}. We then have
\begin{gather*}
\int_{-\infty}^{\infty} \!\! dk (e^{i k |\bm{r} -\bm{r}_j| -i ck (t-t')} - e^{-i k |\bm{r} -\bm{r}_j| -i ck (t-t')} ) \\  
= \frac{2\pi}{c} \delta(t'-t + |\bm{r} -\bm{r}_j|/c) + \frac{2\pi}{c} \delta(t'-t - |\bm{r} -\bm{r}_j|/c) . 
\end{gather*} 
Upon substitution into Eq.~(\ref{eq:Ertplw}) the second term is always zero, and we finally obtain
\begin{equation}
E(\bm{r},t) = \frac{\wp_{eg} k_e^2}{\sqrt{3} \, 2\pi \eps_0} 
\sum_j \frac{e^{-i \om_e (t-|\bm{r} -\bm{r}_j|/c)}}{|\bm{r} -\bm{r}_j|} b_j(t - |\bm{r} -\bm{r}_j|/c) . 
\end{equation}
For a single atom at the origin,  $\bm{r}_j =0$, we have an isotropic spherical wave 
\[
E(\bm{r},t) = \frac{\wp_{eg} k_e^2}{\sqrt{3} \, 2\pi \eps_0} \frac{e^{i (k_e r - \om_e t)}}{r} b(t-r/c) ,
\] 
while the intensity of the emitted radiations at position $\bm{r}$ and time $t$ is given by 
$I(\bm{r},t) = \frac{\eps_0 c}{2} |E(\bm{r},t)|^2 = \frac{\hbar \omega_e}{4 \pi r^2} \frac{1}{2} \Gamma |b(t-r/c)|^2$.

For an ensemble of $N \gg 1$ atoms with density $\rho(\bm{r})$, such that $\int \! d^3 r \rho(\bm{r}) = N$, 
the emitted field is then given by Eq.~(\ref{eq:Ertex}) of the main text.

\section{Single-atom emission into the Gaussian modes}
\label{app:satphot}

Consider the photon emission by a single atom placed at the origin, $\bm{r}_j =0$, 
and therefore $g_k(0) = i \frac{\wp_{eg}}{\hbar} \sqrt{\frac{\hbar \omega_k}{2 \eps_0 LA}}$.  
In Eq.~(\ref{eq:nt}), replacing the summation over the modes by an integration according to 
$\sum_k \to 2 \frac{L}{2\pi c} \int_0^{\infty} d \omega_k$ (factor of 2 is for the two orthogonal 
polarizations of the photon), and substituting $a_k(t)$ with a single atom contributing, 
$b_j \to b$, we have   
\begin{eqnarray}
n(t) &=&  \frac{|\wp_{eg}|^2}{2 \pi \hbar \eps_0 c A} \int_0^{\infty} \!\! d \omega_k \, \omega_k 
\int_0^t \!\! dt' \, b(t') e^{i (\omega_k - \omega_e) t'} 
\nonumber \\ & & \qquad \qquad \qquad  \times
\int_0^t \!\! dt'' \, b^*(t'') e^{-i (\omega_k - \omega_e) t''}  . \quad \label{eq:nt1at}
\end{eqnarray}
We can proceed using the Weisskopf-Wigner approximation as follows 
\begin{eqnarray}
n(t) &=&  \frac{|\wp_{eg}|^2}{2 \pi \hbar \eps_0 c A} 
\iint_0^t \!\! dt' dt'' \, b(t') b^*(t'')  
\int_0^{\infty} \!\!\! d \omega_k \, \omega_k e^{i (\omega_k - \omega_e)(t' - t'')}
\nonumber \\
 & \simeq & \frac{|\wp_{eg}|^2 \omega_e}{2 \pi \hbar \eps_0 c A} 
\iint_0^t \!\! dt' dt'' \, b(t') b^*(t'')  
\int_{-\omega_e}^{\infty} \!\!\! d \omega_k' \, e^{i \omega_k' (t' - t'')}
\nonumber \\
 & \simeq & \frac{|\wp_{eg}|^2 \omega_e}{2 \pi \hbar \eps_0 c A} 
\iint_0^t \!\! dt' dt'' \, b(t') b^*(t'')  
\int_{-\infty}^{\infty} \!\!\! d \omega_k' \, e^{i \omega_k' (t' - t'')}
\nonumber \\
 & = &  \frac{|\wp_{eg}|^2 \omega_e}{\hbar \eps_0 c A} 
\iint_0^t \!\! dt' dt'' \, b(t') b^*(t'') \, \delta (t'-t'') 
\nonumber \\
&=& \frac{|\wp_{eg}|^2 \omega_e}{\hbar \eps_0 c A} 
\int_0^t \!\! dt' \, |b(t')|^2 
\nonumber \\
&=& \frac{\varsigma}{2A} \Gamma \int_0^t dt' \, |b(t')|^2 ,
\end{eqnarray}
where $\varsigma = 3 \lambda_e^2/2\pi$ is the resonant absorption cross section of the atom.
This is a very simple and intuitive result: For an initially excited atom, $|b(t')|^2 = e^{-\Gamma t'}$, 
we have that $n(t \to \infty) =  \varsigma/(\pi w_0^2)$, i.e., the amount of radiation collected
by the Gaussian paraxial optics is proportional to the ratio of the atomic absorption cross-section 
to the cross-section of the focused Gaussian beam at the atomic position.

\section*{Declarations}

\subsection*{Availability of data and materials}
Not applicable.

\subsection*{Competing interest}
The authors declare that they have no competing interests.

\subsection*{Funding}
The work of A.K., P.D. and A.V. was supported by the National Research, Development and Innovation Office of Hungary 
(Project Nos. K115624, 2017-1.2.1-NKP-2017-00001). 
A.V. was also supported by the J\' anos Bolyai Research Scholarship of the Hungarian Academy of Sciences. 
T.B. and N.T.Z. acknowledge support from the Carlsberg Foundation and the Independent Research Fund Denmark.
J.F. and D.P. were supported by the DFG SPP 1929 GiRyd and DFG Project No. 394243350.
D.P. was also supported by the EU QuanERA Project PACE-IN and by 
the Alexander von Humboldt Foundation in the framework of the Research Group Linkage Programme.

\subsection*{Author’ contributions}
D.P., N.T.Z., and J.F. initiated the work. All authors discussed the approach to and methodology of the problem. 
A.K and T.B. performed the calculations under supervision of A.V., P.D., and D.P. 
A.K., A.V., P.D., and D.P. wrote the manuscript. A.K. prepared the datasets, A.V. and D.P. prepared the figures. 
All authors discussed and approved the final version of the manuscript.

\end{document}